\documentclass[a4paper]{jpconf}

\usepackage{amssymb}
\usepackage{graphicx}

\newcommand{\trace}{\text{Tr}\:}
\newcommand{\text}[1]{\mbox{\rm #1}}
\newcommand{\cfrac}[2]{{\displaystyle\frac{#1}{#2}}
\kern-\nulldelimiterspace}
\newcommand{\gsm}[3]{\left(\mathbf{#1},\mathbf{#2}\right)_{\,#3}}
\newcommand{\gsmbar}[3]{\left(\mathbf{\bar{#1}},\mathbf{#2}\right)_{\,#3}}

\begin{document}

\title{Unifying gauge couplings at the string scale\footnote{Talk
given by D. Emmanuel-Costa on the RTN meeting ``The Quest for
Unification: Theory Confronts Experiment'', 11 - 18
September 2005, Corfu, Greece.}}

\author{David Emmanuel-Costa and Ricardo Gonz\'alez Felipe}

\address{ Departamento de F\'{\i}sica and Centro de F\'{\i}sica
Te\'orica de Part\'{\i}culas\\ Instituto Superior T\'ecnico,
Av. Rovisco Pais, P-1049-001 Lisboa, Portugal}

\ead{david.costa@ist.utl.pt, gonzalez@cftp.ist.utl.pt}

\begin{abstract}
Using the current precision electroweak data, we look for the minimal
particle content which is necessary to add to the standard model in
order to have a complete unification of gauge couplings and gravity at
the weakly coupled heterotic string scale.  We find that the addition
of a vector-like fermion at an intermediate scale and a non-standard
hypercharge normalization are in general sufficient to achieve this
goal at two-loop level. Requiring the extra matter scale to be below
the TeV scale, it is found that the addition of three vector-like
fermion doublets with a mass around 700~GeV yields a perfect
string-scale unification, provided that the affine levels are
$(k_Y\,,k_2\,,k_3)=(13/3\,,1\,,2)\,$, as in the $SU(5) \times SU(5)$
string-GUT.  Furthermore, if supersymmetry is broken at the
unification scale, the Higgs mass is predicted in the range 125 GeV -
170 GeV, depending on the precise values of the top quark mass and
$\tan \beta$ parameter.
\end{abstract}

\section{Introduction}

\begin{figure}[t]
\caption{\label{fig:g} Gauge coupling running in the MSSM.}
 \centering\includegraphics[width=12cm]{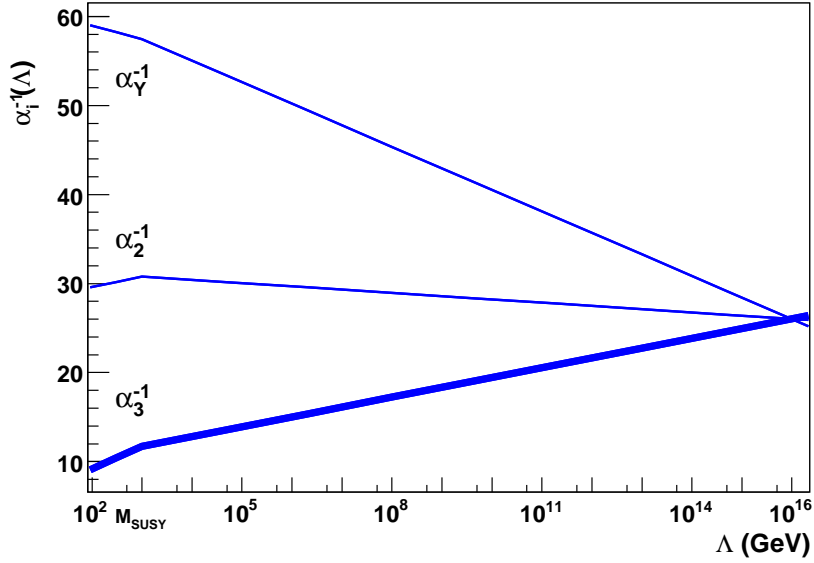}
\end{figure}

\begin{figure}[t]
\caption{\label{fig:g1} Gauge coupling running in the SM with $k_Y=13/10$.}
 \centering\includegraphics[width=12cm]{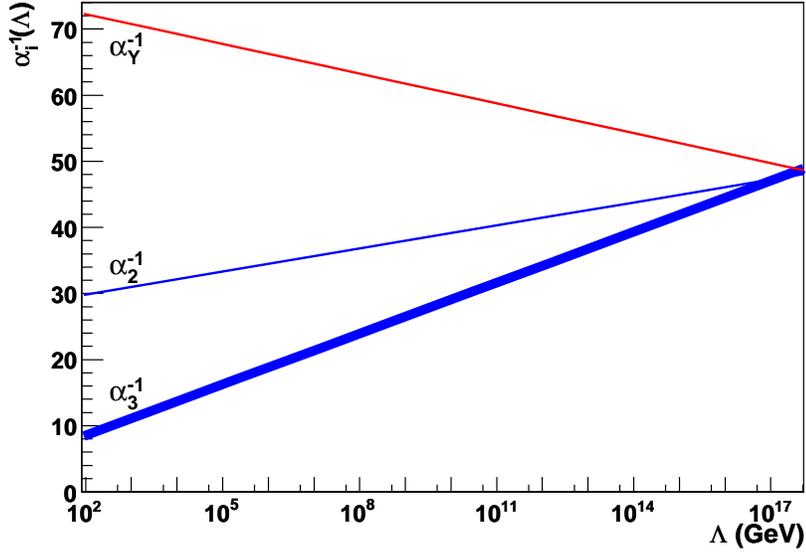}
\end{figure}

Unification of gauge couplings has always been one of the few solid
pieces of evidence in favor of supersymmetry. It is well known that
the extrapolation of low-energy data within the framework of the MSSM
yields an almost perfect unification of gauge couplings at the scale
$\Lambda_{MSSM} \approx 2 \times 10^{16}$~GeV (see
Figure~\ref{fig:g}), which is lower than the typical string scale,
$\Lambda_S\gtrsim 10^{17}$ GeV. The resolution of this discrepancy has
been the subject of many studies and several paths to unification have
been proposed~\cite{Dienes:1996du,Munoz:2001yj,Dienes:1995bx}. On the
other hand, it is remarkable that, in the non-supersymmetric SM, the
one-loop $g_2$ and $g_3$ gauge couplings already unify at a scale
$\Lambda_{SM} \approx 10^{17}$~GeV (see Figure~\ref{fig:g1}), which is
close to the unification scale predicted by the string theory. In this
case, gauge coupling unification could be achieved for a hypercharge
normalization $k_Y \approx 13/10$~\cite{Dienes:1996du}. However, if
two-loop effects are taken into account, the above scale should be at
most $\Lambda_{SM} \approx 4 \times 10^{16}$~GeV~\cite{Cho:1997gm},
which is one order of magnitude smaller than the expected string
scale. For high-scale supersymmetry breaking, it has been recently
shown that gauge coupling unification can be achieved at about $2
\times 10^{16}$~GeV in axion models with SM vector-like
fermions~\cite{Barger:2004sf}, or at $10^{16-17}$~GeV in the SM with
suitable normalizations of the $U(1)_Y$, which can be realized in
specific orbifold GUTs~\cite{Barger:2005gn}. Nevertheless, the
unification scale in all of these cases is somehow below the expected
string scale.

The phenomenology of $E_8\times E_8$ heterotic string
theory~\cite{Gross:1985fr} exhibits many of the attractive features of
the low-energy physics that we see today. In particular, the
four-dimensional standard model (SM) gauge group $G_{SM}=SU(3)_C
\times SU(2)_L \times U(1)_Y$ and its generations can be easily
incorporated. String theory also offers an elegant explanation for the
doublet-triplet splitting problem~\cite{Pokorski:1997mv}. Moreover,
the unification of gauge couplings and gravity is an intrinsic
property of heterotic string theory. Remarkably, unification of
couplings is a prediction of string theories even without any grand
unified theory (GUT) below the Planck scale. Indeed, gauge and
gravitational couplings unify at tree level as~\cite{Ginsparg:1987ee}
\begin{equation} \label{unif_rel}
\alpha_{\rm string}=\frac{2\, G_N}{\alpha^\prime}=k_i\,\alpha_i \,,
\end{equation}
where $\alpha_{\rm string}=g_{\rm string}^2/4\pi$ is the string-scale
unification coupling constant, $G_N$ is the Newton constant,
$\alpha^\prime$ is the Regge slope, $\alpha_i=g_i^2/4\pi\, (i=Y,2,3)$
are the gauge couplings and $k_i$ are the so-called affine or
Ka\v{c}-Moody levels at which the group factors $U(1)_Y$, $SU(2)_L$
and $SU(3)_C$ are realized in the four-dimensional string. The
appearance of non-standard affine levels $k_i$ plays an important role
in string theories. While the non-Abelian factors $k_2$ and $k_3$
should be positive integers, the Abelian factor $k_Y$ can take a
priori any arbitrary value, only constrained to be $k_Y>1$ for the
right-handed electron to have a consistent hypercharge
assignment. Furthermore, these factors determine the value of the
mixing angle $\sin\theta_W$ at the string scale.  

It may be possible that the string compactifies in four dimensions not
to the SM group, but to a simple group which acts as a unified
group. In this case $\Lambda_{\text{\tiny GUT}}=\Lambda_S$ and the
Ka\v{c}-Moody levels are fixed by the group structure,
\begin{equation}
k_i\,=\,\frac{\trace T^2}{\trace T^2_i}\,,
\end{equation}
 where $T$ is a generator of the subgroup $G_i$ properly normalized
over a representation $R$ of the string-GUT group and $T_i$ is the
same generator but normalized over the representation of the subgroup
embedded into $R$ \cite{Perez-Lorenzana:1999tf}. For illustration, the
Ka\v{c}-Moody levels for different possible string-GUT models are
presented in Table \ref{tab:1}.

%%%%%%%%%%%%%%%%%%%%%%%%
\begin{table}
\caption{\label{tab:1} Ka\v{c}-Moody levels for several possible string-GUT
models; $F=1,2,\dots$ stands for the number of families in that
particular model \cite{Perez-Lorenzana:1999tf}. \\}
\begin{center}
\begin{tabular}{lccc}
\hline
{\bf Group} & {\bf $k_Y$} & 
{\bf $k_2$} & {\bf $k_3$}\\
\hline
{\small
  $\!\!\!\left.\begin{array}{l}
SU(5),\, SO(10),\,SU(15),\\\
E_6,\,E_8,\,\left[SU(3)\right]^3\times Z_3,\\
SU(16),\,SU(8)\times~SU(8),\,SO(18)
\end{array}
\right\}$ \textbf{Canonical}}
& 5/3 & 1 & 1\\[7mm]
$\left[SU(3)\right]^4\times Z_4$ & 5/3 & 1 & 2\\[3mm]
$SU(5)\times SU(5),\,SO(10)\times SO(10)$ & 13/3 & 1 & 2\\[3mm]
$\left[SU(6)\right]^3\times Z_3$ & 14/3 & 3 & 1\\[3mm]
$\left[SU(6)\right]^4\times Z_4$ & 19/3 & 3 & 2\\[3mm]
$E_7$ & 2/3 & 2 & 1\\[3mm]
$\left[SU(4)\right]^3\times Z_3$ & 11/3 & 1 & 1\\[3mm]
$\left[SU(2\mathrm{F})\right]^3\times Z_4$ & (6F-4)/3 & F & 1\\[3mm]
$\left[SU(2\mathrm{F})\right]^4\times Z_4$ & (9F-8)/3 & F & 2\\[3mm]
\hline
\end{tabular}
\end{center}
\end{table}
%%%%%%%%%%%%%%%%%

Since string theory relates a dimensionless gauge coupling to a dimensionful
gravitational coupling, Equation~(\ref{unif_rel}) itself predicts the unification
scale $\Lambda = g_{\rm string} M_P$, where $M_P = 1.22 \times 10^{19}$~GeV is
the Planck mass. This scale is lowered by the inclusion of one-loop string
effects and in the weak coupling limit one finds~\cite{Kaplunovsky:1987rp}
\begin{equation} \label{Lambda}
    \Lambda = g_{\rm string} \Lambda_S\,,
\end{equation}
where $\Lambda_S$ is given by
\begin{equation} \label{Lambda_s}
\Lambda_S\,=\,\frac{e^{(1 - \gamma)/2}\,3^{-3/4}}{4\pi}\,M_P \,\approx
\,5.27\times10^{17}\,\,\text{GeV}\,,
\end{equation}
$\gamma \approx 0.577$ is the Euler constant. It has also been
noted~\cite{Witten:1996mz,Banks:1996ss} that  the unification scale in the
strong coupling limit can be much lower than the perturbative result given by
Equation~(\ref{Lambda}). Yet, it is not clear whether unification is a robust
prediction in this case.

Next, using the current precision electroweak data, we study the
problem of gauge coupling unification within string theory, with the
aim to look for the minimal particle content which is necessary to add
to the SM in order to achieve unification at the weakly coupled
heterotic string scale \cite{Emmanuel-Costa:2005nh}.

\section{One-loop analysis}

The evolution of the gauge coupling constants at one loop is governed by the
renormalization group equations (RGE)
\begin{equation}
\label{eq:sol}
 \alpha_i^{-1}(\mu)\;=\;\alpha_{i\,Z}^{-1}\,-\,\frac{b_i}{2\pi}\,
\log\frac{\mu}{M_Z}\ ,
\end{equation}
where $\alpha_{i\,Z} \equiv \alpha_i(M_Z)$ and the $\beta$-function
coefficients $b_i$ are given by
\begin{equation}
b_i=\frac13\sum_R\left[\:s(R)\,N_i(R)\:\right]-\frac{11}3\,C_2(G_i)\ ,
\end{equation}
for non-supersymmetric theories. The function $s(R)$ is 1 for complex scalars,
2 for chiral fermions and 4 for vector-like fermions. The Casimir group
invariant for the adjoint representation, $C_2(G_n)$, is $n$ for $SU(n)$ groups
and null for a $U(1)$ group. The functions $N_i(R)$ encode the group structure
contributions as follows
\begin{equation}
 N_i(R)\,=\,T_i(R)\prod_{j\neq i}d_j(R)\ ,
\end{equation}
where $d_i(R)$ is the dimension of the representation concerning the invariant
subgroup $G_i$ and $T_i(R)$ is the Dynkin index which, in our convention, is
$1/2$ for the fundamental representations of $SU(n)$ groups and $y^2$ for the
$U(1)_{Y}$ group. We use the convention that the hypercharge $Y=Q-T_{3L}$. In
particular, for the SM with $N$ generations and $n_H$ complex Higgs doublets
one finds
\begin{equation} \label{bSM}
b_Y = \frac{20}9\,N+\frac{n_H}{6}\,,\quad b_2= \frac43\,N+
\frac{n_H}{6}-\frac{22}{3}\,,\quad b_3= \frac43\,N-11\ .
\end{equation}

Let us now examine the one-loop running of the gauge couplings. The unified
coupling constant $\alpha_{\rm string}$ at the scale $\Lambda$ is expressed in
terms of the $SU(3)_{C}$, $SU(2)_{L}$ and $U(1)_{Y}$ gauge couplings and the
corresponding affine levels $k_i$ through
Equations~(\ref{unif_rel})-(\ref{Lambda_s}). Thus, at the unification scale
$\Lambda$, Equation~(\ref{eq:sol}) implies
\begin{equation}
\label{eq:sol1}
 \alpha_{i\,Z}^{-1}\;=\;k_i\,\alpha_{\rm string}^{-1}
\,+\,\frac{b_i}{2\pi}\,\log\frac{\Lambda}{M_Z}\ ,
\end{equation}
with the additional constraint
\begin{equation}
\alpha_{\rm string}\,=\,\frac1{4\pi}\,\left(\frac{\Lambda}{\Lambda_S}\right)^2\
,
\end{equation}
which reflects the stringy nature of the unification.

These equations can be analytically solved to determine the scale $\Lambda$. We
obtain
\begin{equation} \label{Lambda_SM}
\left(\frac{\Lambda_S}{\Lambda}\right)^2\,=\,-\frac{b_i}{16\pi^2\,k_i}\;
W_{-1}\left[-\left(\cfrac{4\pi\,\Lambda_S}{M_Z}\right)^2\cfrac{k_i}{b_i}
\;e^{-4\pi/(b_i\, \alpha_{iZ})\,}\right]\ ,
\end{equation}
where $W_{-1}(x)$ is the $k=-1$ real branch of the Lambert $W_k$
function~\cite{Corless:1996}.

In our numerical calculations we shall use the following electroweak input data
at the  $Z$ boson mass scale $M_Z \simeq
91.2$~GeV~\cite{Eidelman:2004wy,Bethke:2004uy}:
\begin{eqnarray}
\label{n:MZ}
 \alpha^{-1}(M_Z) &=& 128.91 \pm 0.02\,,\nonumber\\
 \sin^2\theta_W(M_Z) &=& 0.23120 \pm  0.00015\,,\\
 \alpha_s(M_Z) &=& 0.1182 \pm 0.0027\,,\nonumber
\end{eqnarray}
for the fine structure constant $\alpha$, the weak mixing angle $\theta_W$ and
the strong coupling constant $\alpha_s$, respectively. The top quark pole mass
$M_t^{\rm pole}$ is taken as~\cite{Azzi:2004rc}
\begin{equation}
\label{Mtpole}
 M_t^{\rm pole}\,=\, 178.0 \pm 4.3~\text{GeV}\,,
\end{equation}
and the Higgs vacuum expectation value $v=174.1$~GeV.

Using the SM coefficients $b_i$ given in Equation~(\ref{bSM}) and assuming $k_2 =
k_3=1$, we obtain from Equation~(\ref{Lambda_SM}), $\Lambda \approx 2.7 \times
10^{17}$~GeV, which in turn implies $\alpha_{\rm string} =0.021$. Substituting
these values into Equation~(\ref{eq:sol1}) we find $\alpha_s(M_Z) \simeq 0.1239$, a
value which is clearly outside the experimental range given in
Equation~(\ref{n:MZ}). The above result already indicates that the string-scale
unification of gauge couplings requires either non-perturbative (or
higher-order perturbative) string effects to lower the unification scale or
extra matter particles to modify the RGE evolution of the gauge couplings. It
is precisely the second possibility that we will consider here.

Anticipating a possible string-GUT compactification scenario, we shall restrict
our analysis to the inclusion of fermions in real irreducible representations.
The addition of chiral fermions leads in general to anomalies and their masses
are associated to the electroweak symmetry breaking, which imposes further
constraints. Also, the introduction of new light scalars requires additional
fine-tunings. Thus, we shall consider the following fermionic
states~\cite{Giudice:2004tc}:
\begin{equation} \label{newpart}
\begin{array}{l@{\quad}l}
Q\,=\,\gsm{3}{2}{1/6}+\gsmbar{3}{2}{\text{-}1/6}\,, &
L\,=\,\gsm{1}{2}{\text{-}1/2}+\gsm{1}{2}{1/2}\,,\\[2mm]
U\,=\,\gsm{3}{1}{2/3}+\gsmbar{3}{1}{\text{-}2/3}\,,&
D\,=\,\gsm{3}{1}{\text{-}1/3}+\gsmbar{3}{1}{1/3}\,,\\[2mm]
E\,=\,\gsm{1}{1}{\text{-}1}+\gsm{1}{1}{1}\,,&
X\,=\,\gsm{3}{2}{\text{-}5/6}+\gsmbar{3}{2}{5/6}\,,\\[2mm]
G\,=\,\gsm{8}{1}{0}\,,& V\,=\,\gsm{1}{3}{0}\,.
\end{array}
\end{equation}
These can naturally appear in extensions of the SM as a part of some incomplete
GUT multiplets. They are present, for instance, in the $\mathbf{5+\bar{5}}$,
$\mathbf{10+\overline{10}}$ and $\mathbf{24}$ irreducible representations of
$SU(5)$. The addition of such matter states gives corrections to the $b_i$
coefficients in the gauge coupling running. Denoting by $\Delta_i$ these
corrections, one has
\begin{eqnarray}
\Delta_Y & = & \frac29\,n_Q+\frac23\,n_L
       +\frac{16}9\,n_U+\frac49\,n_D+\frac43\,n_E+\frac{50}9\,n_X\ ,\\[4mm]
\Delta_2 & = & 2\,n_Q+\frac23\,n_L+2\,n_X
       +\frac43\,n_V\ ,\\[4mm]
\Delta_3 & = & \frac43\,n_Q+\frac23\,n_U+\frac23\,n_D+\frac43\,n_X+2\,n_G\ ,
\end{eqnarray}
where $n_r$ denotes the number of multiplets belonging to the irreducible
representations $r$ given in Equation~(\ref{newpart}). The string unification
conditions (\ref{eq:sol1}) also get modified,
\begin{equation}
\label{eq:sol2}
 \alpha_{i\,Z}^{-1}\;=\;k_i\,\alpha_{\rm string}^{-1}
\,+\,\frac{b_i}{2\pi}\,\log\frac{\Lambda}{M_Z}
\,+\,\frac{\Delta_i}{2\pi}\,\log\frac{\Lambda}{M} \ ,
\end{equation}
where $M$ is the new-physics threshold. Notice that we assume a common mass
scale for the extra matter content, once we are interested in minimal scenarios
which could lead to a successful unification. The solution of the above
equations is now given by
\begin{equation} \label{Lambda1loop}
\left(\frac{\Lambda_S}{\Lambda}\right)^2\,=\, -\frac{1}{16\pi^2\,\rho}\;
W_{-1}\!\!\left[-\left(\cfrac{4\pi\,\Lambda_S}{M_Z}\right)^2
\rho\,e^{-4\pi\,\eta}\:\right]\ ,
\end{equation}
for the unification scale and
\begin{equation}\label{M1loop}
\frac{M}{M_Z}\,=\,\left(\frac{\Lambda}{M_Z}\right)^{\rho^\prime-1}\,e^{-2\pi
\eta^\prime}\,,
\end{equation}
for the threshold, where
\begin{eqnarray}
\rho\,&\equiv&\,
\frac{\Delta_3\,k_2-\Delta_2\,k_3}{\Delta_3\,b_2-\Delta_2\,b_3}\ ,\quad
\eta\,\equiv\,
\frac{\Delta_3\,\alpha^{-1}_{2\,Z}-\Delta_2\,\alpha^{-1}_{3\,Z}}{
\Delta_3\,b_2-\Delta_2\,b_3}\ ,\nonumber\\ \\
 \rho^\prime\,&\equiv&\,
\frac{b_3\,k_2-b_2\,k_3}{\Delta_3\,k_2-\Delta_2\,k_3}\ ,\quad
\eta^\prime\,\equiv\, \frac{k_3\,\alpha^{-1}_{2\,Z}-k_2\,\alpha^{-1}_{3\,Z}}{
k_3\,\Delta_2-k_2\,\Delta_3}\nonumber \ .
\end{eqnarray}

Finally, the hypercharge normalization $k_Y$ is determined from
Equation~(\ref{eq:sol2}):
\begin{equation} \label{kY1loop}
k_Y\,=\,\alpha_{\rm string} \left[\:\alpha^{-1}_{1\,Z}
\,-\,\frac{b_Y}{2\pi}\log\frac{M}{M_Z}
\,-\,\frac{\Delta_Y}{2\pi}\log\frac{\Lambda}{M} \:\right]\ .
\end{equation}

Using Equations~(\ref{Lambda1loop})-(\ref{kY1loop}), it is straightforward to obtain
all the possible solutions that lead to the string-scale unification of
couplings at one-loop order. Here we present only those which are minimal, i.e.
those which require the addition of a single extra particle with a mass scale
$M$. The results are given in Table~\ref{tab1}. There exist 3 minimal
solutions, namely, $n_U=1\,$, $n_D=1\,$ and $n_G=1$, which correspond to the
addition of an up-type or down-type vector-like fermion or one gluino-type
fermion, respectively, with quantum numbers as given in Equations~(\ref{newpart}).
In all three cases the presence of a non-canonical hypercharge normalization,
$k_Y \neq 5/3$, is required. We have taken the non-Abelian affine levels $k_2$
and $k_3$ to be equal to 1 or 2, which are the preferred values from the
string-model building viewpoint~\cite{Dienes:1996du}. We also notice that no
minimal solution was found with $k_2 \neq k_3$.

\begin{table}[ht]
\caption{\label{tab1}Minimal extra matter content which leads to string-scale
unification at one loop. The results for the new-physics threshold $M$, the
unification scale $\Lambda$ and the hypercharge affine level $k_Y$ are
presented for the central values given in Equation~(\ref{n:MZ}).}

\begin{center}
\scalebox{0.95}{
  \begin{tabular}{cc@{\ \  }c@{\ \ \ \ }c@{\ \ }c@{\ \ \ \ }c@{\ \  }c}
  \\\hline
   & \multicolumn{2}{c}{$n_U=1$} & \multicolumn{2}{c}{$n_D=1$} &
     \multicolumn{2}{c}{$n_G=1$} \\
     & $k_{2,3}=1$ & $k_{2,3}=2$ &
       $k_{2,3}=1$ & $k_{2,3}=2$ &
       $k_{2,3}=1$ & $k_{2,3}=2$ \\[1mm]
       \hline 
       $M\,(\text{GeV})$ & $6.8\times 10^{15}$ & $1.3\times 10^{15}$ &
       $6.8\times 10^{15}$ & $1.3\times 10^{15}$ &
       $7.9\times 10^{16}$ & $5.8\times 10^{16}$ \\
       $\Lambda\,(\text{GeV})$ & $2.7\times 10^{17}$ & $3.8\times 10^{17}$ &
       $2.7\times 10^{17}$ & $3.8\times 10^{17}$ &
       $2.7\times 10^{17}$ & $3.8\times 10^{17}$ \\
       $k_Y$ & $1.24$  & $2.44$  & $1.26$ & $2.49$ & $1.26$ & $2.50$\\
       \hline\\
  \end{tabular} }
  \end{center}
  \end{table}

\begin{table}[ht]
\caption{\label{tab2}Minimal solutions which lead to string-scale unification
at two-loop order. We use the central values for the electroweak input data
given in Equations~(\ref{n:MZ}) and (\ref{Mtpole}).}

\begin{center}
\scalebox{0.95}{
  \begin{tabular}{cc@{\ \  }c@{\ \ \ \ }c@{\ \ }c@{\ \ \ \ }c@{\ \  }c}
\\\hline
   & \multicolumn{2}{c}{$n_U=1$} & \multicolumn{2}{c}{$n_D=1$} &
     \multicolumn{2}{c}{$n_G=1$} \\
     & {\small $k_{2,3}\!=\!1$} & {\small $k_{2,3}\!=\!2$} &
       {\small $k_{2,3}\!=\!1$} & {\small $k_{2,3}\!=\!2$} &
       {\small $k_{2,3}\!=\!1$} & {\small $k_{2,3}\!=\!2$} \\
\hline $M\,(\text{GeV})$ & $7.2\times 10^{12}$ & $1.5\times 10^{12}$ &
                    $7.1\times 10^{12}$ & $1.4\times 10^{12}$ &
                    $8.2\times 10^{15}$ & $6.1\times 10^{15}$ \\
$\Lambda\,(\text{GeV})$ & $2.7\times 10^{17}$ & $3.8\times 10^{17}$ &
                          $2.7\times 10^{17}$ & $3.8\times 10^{17}$ &
                          $2.7\times 10^{17}$ & $3.8\times 10^{17}$ \\
$k_Y$ & $1.20$  & $2.35$  & $1.25$ & $2.47$ & $1.26$ & $2.50$\\
\hline\\
\end{tabular}
}
\end{center}
\end{table}

\section{Two-loop gauge coupling unification}

To perform a more precise analysis of string unification, a two-loop RGE study
becomes necessary. We make use of the two-loop RGEs of gauge
couplings~\cite{Machacek:1983tz}, which include the one-loop Yukawa coupling
running and take properly into account the new physics contributions and
threshold. In Table~\ref{tab2} we present the two-loop results for the minimal
one-loop solutions given in Table~\ref{tab1}. As in the one-loop case, no
solution was found with $k_2\neq k_3$.

\begin{table}[ht]
\caption{\label{tab3}Minimal extra particle content with a mass below the TeV
scale, which leads to unification at two-loop order. We use the electroweak
input data given in Equations~(\ref{n:MZ}) and (\ref{Mtpole}). The non-Abelian
affine levels are $k_2=1$ and $k_3=2$ in all cases. The quantities in brackets
reflect the effects of the $\alpha_s(M_Z)$ uncertainty.}

\begin{center}
\scalebox{1.0}{
  \begin{tabular}{c@{\ \ \ \ }ccc}
\\\hline
& $M\,(\text{GeV})$ & $\Lambda\,(\text{GeV})$ & $k_Y$
\\\hline
$n_Q\,=\,3$ & $\left[\,653\,,\,823\,\right]$ & $5.2\times 10^{17}$ &
$\left[\,4.27\,,\,4.37\,\right]$ \\[1mm]
$n_Q\,=\,2\,, n_X\,=\,1$ & $\left[\,676\,,\,852\,\right]$ & $5.2\times 10^{17}$
& $\left[\,1.98\,,\,2.00\,\right]$ \\[1mm]
$n_Q\,=\,2\,, n_V\,=\,1$ & $\left[\,459\,,\,587\,\right]$ & $4.6\times 10^{17}$
& $\left[\,3.37\,,\,3.42\,\right]$ \\[1mm]
$n_Q\,=\,1\,, n_X\,=\,1\,, n_V\,=\,1$ & $\left[\,475\,,\,607\,\right]$ &
$4.6\times 10^{17}$ & $\left[\,1.60\,,\,1.61\,\right]$ \\[1mm]
$n_Q\,=\,1\,, n_V\,=\,2$ & $\left[\,351\,,\,452\,\right]$ &
$4.1\times 10^{17}$ & $\left[\,2.81\,,\,2.84\,\right]$ \\[1mm]
$n_X\,=\,1\,, n_V\,=\,2$ & $\left[\,363\,,\,468\,\right]$ &
$4.1\times 10^{17}$ & $\left[\,1.37\,,\,1.37\,\right]$ \\[1mm]
$n_V\,=\,3$ & $\left[\,283\,,\,367\,\right]$ & $3.8\times 10^{17}$ &
$\left[\,2.43\,,\,2.44\,\right]$ \\[1mm]
\hline\\
\end{tabular}
}
\end{center}
\end{table}

It turns out that the unification scale $\Lambda$ and the hypercharge
normalization are not very sensitive to higher order corrections. This can be
readily seen by comparing the one-loop results of Equations~(\ref{Lambda1loop}) and
(\ref{kY1loop}) with the two-loop values numerically obtained (see
Table~\ref{tab2}). On the other hand, the new-physics threshold $M$ can be
significantly altered by such corrections. In particular, we notice that while
at one loop the solutions $n_U=1$ and $n_D=1$ require an intermediate scale of
the order of $10^{15}-10^{16}$~GeV, this scale is lowered to
$10^{12}-10^{13}$~GeV at two-loop order.  One may ask whether such an
intermediate mass scale could be naturally generated. In principle, it might be
due to the possible presence of nonrenormalizable higher-order operators or
could be associated with an approximate global symmetry, such as a chiral
symmetry of Peccei-Quinn type.

We have also searched for minimal solutions where the new matter states have a
mass scale below the TeV scale. Seven solutions were found, which are listed in
Table~\ref{tab3}. All of them require the non-Abelian affine levels to be
$k_2=1$ and $k_3=2$. Of particular interest is the first solution with three
vector-like fermion doublets, i.e. $n_Q=3$. Not only it yields a perfect
string-scale unification at $g_{\rm string} \approx 1$, but also, for
$\alpha_s(M_Z) = 0.119$ and $M = 710$~GeV, it implies the hypercharge
normalization $k_Y =13/3\,$, thus suggesting an $SU(5) \times SU(5)$ or
$SO(10) \times SO(10)$ string-GUT
compactification~\cite{Davidson:1987mi,Perez-Lorenzana:1999tf}.

\section{Higgs boson mass}

In the string landscape~\cite{Bousso:2000xa}, the supersymmetry breaking scale
can be high and the SM (with, eventually, some residual matter content) is the
simplest effective theory all the way down to low energies. In this scenario,
the mass of the yet undiscovered Higgs boson appears to be the most relevant
parameter. In general, supersymmetric models contain one pair of Higgs doublets
$H_u$ and $H_d\,$. The combination $\phi \equiv \sin\beta\,H_u-\cos\beta\,i
\sigma_2 H^{\ast}_d\,$ is typically chosen as the fine-tuned SM Higgs doublet
$\phi$ with a small mass term. If supersymmetry is broken at the string scale,
the Higgs boson quartic coupling $\lambda$ at the unification scale is then
given by
\begin{equation}
  \lambda(\Lambda)\,=\,\frac{1}{4} \left[\,g^2(\Lambda)\,+\,
  g^{\prime\,2}(\Lambda)\,\right] \cos^2 2\beta
  = \,\pi\,\alpha_{\rm string} \left(\,\frac{1}{k_Y}\,+\,\frac{1}{k_2}
  \,\right)\cos^2 2\beta\ .
\end{equation}
After evolving this coupling down to the electroweak scale, one can calculate
the Higgs boson mass $m_H$ by minimizing the one-loop effective potential,
\begin{equation}
V\,=\,-m^2\,(\phi^{\dagger}\phi)\, +\,\frac{\lambda}{2}\,(\phi^{\dagger}\phi)^2
\,+\,3\,\alpha_t^2\,(\phi^{\dagger}\phi)^2 \left[\log
\frac{4\pi\,\alpha_t\,(\phi^{\dagger}\phi)}{Q^2} -\frac32\right]\ ,
\end{equation}
which includes top quark radiative corrections. Here $m^2$ is the Higgs mass
parameter, $\alpha_t=y^2_t/4\pi$ is the top quark coupling and the scale $Q$ is
chosen at $Q^2=m_H^2$. The resulting Higgs mass can be written in the following
simple analytical form
\begin{equation}
  m_H^2\,=\,12\,v^2\,\alpha_t^2\,W_0\left(\frac{\pi}{3\,\alpha_t}
\,e^{\frac{\lambda}{6\,\alpha_t^2}}\right)\ ,
\end{equation}
where $W_{0}(x)$ is the principal branch of the Lambert $W$ function.

\begin{figure}[t]
\caption{\label{fig1}The prediction for the Higgs boson mass in the SM extended
with one down-type vector-like fermion. The predicted Higgs mass for the other
two solutions given in Table~\ref{tab2} ($n_U=1$ and $n_G=1$) is similar to the
one depicted in the figure.}
\begin{center}
\includegraphics[scale=0.75]{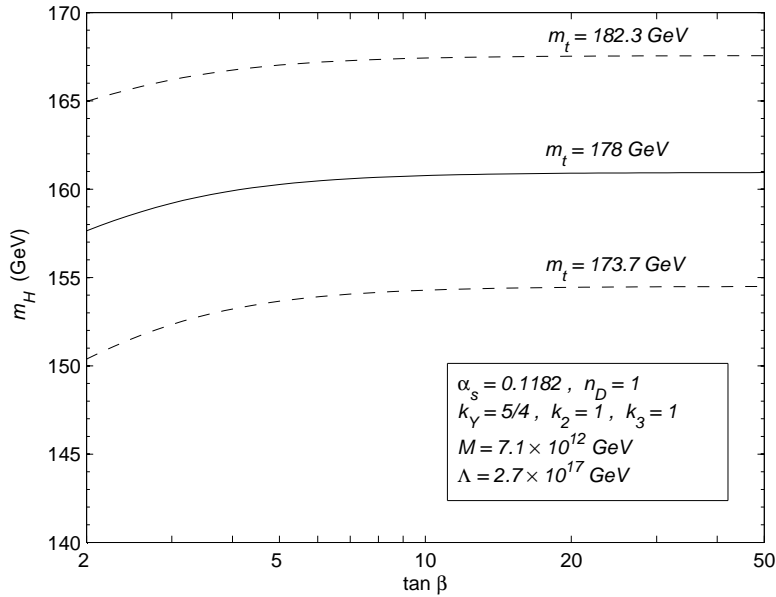}
\end{center}
\end{figure}

\begin{figure}[t]
\caption{\label{fig2}The predicted Higgs boson mass in the SM extended with
three vector-like fermion doublets.}
\begin{center}
\includegraphics[scale=0.75]{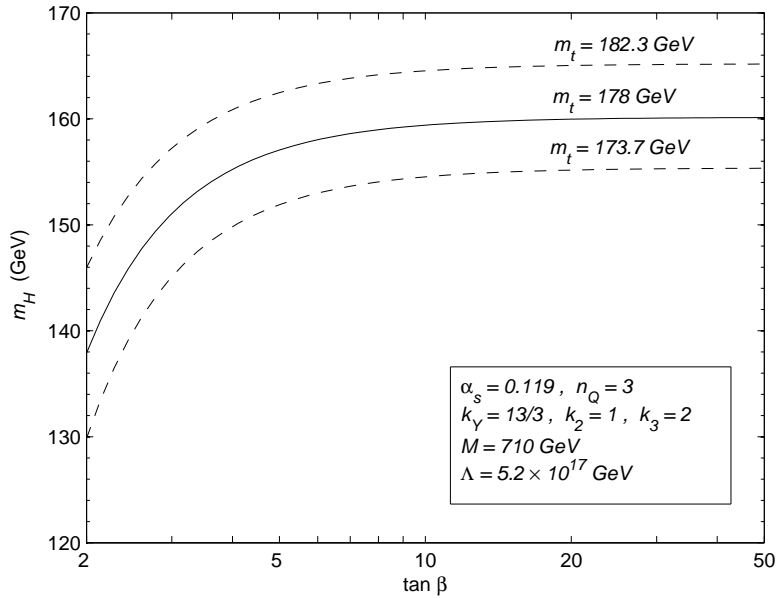}
\end{center}
\end{figure}

The predictions for the Higgs mass are presented in Figs.~\ref{fig1} and
\ref{fig2}, for the minimal string unification solutions found in the previous
section (cf. Tables \ref{tab2} and \ref{tab3}). If we vary $m_t$ within the
$1\sigma$ range given in Equation~(\ref{Mtpole}) and $\tan\beta$ from 2 to 50, the
predicted Higgs boson mass will range from 150~GeV to 167~GeV for the solutions
$n_{U,D,G}=1$, while for the solution $n_Q=3$ the predicted mass varies in the
range from 130~GeV to 165~GeV. If we take into account the presently allowed
$\alpha_s(M_z)$ uncertainty, these intervals are slightly larger and we find
$125~\text{GeV} \lesssim m_H \lesssim 170~\text{GeV}$. Future colliders will
have the potential for the discovery of a Higgs boson with a mass in the above
range~\cite{Buscher:2005re}.

\section{Conclusion}

String theory offers us a consistent framework for the unification of all the
fundamental interactions including gravity. For a weakly coupled heterotic
string, the unification scale is expected around $5\times 10^{17}$~GeV, which
is too high to be achieved in the SM or MSSM, even with a non-canonical
normalization of the hypercharge. A possible way to reconcile the GUT and
string scales is the addition of new matter states to the particle spectrum. In
this talk we have presented some minimal solutions based on the introduction
of vector-like fermions. Working at two-loop order, three minimal solutions
were found, which correspond to the presence at an intermediate scale of an
up-type, down-type or gluino-type fermion with affine levels $k_2=k_3=1$ and
$k_Y \approx 6/5\,, 5/4\,, 63/50\,$, respectively.

Another interesting issue is the existence of new particles with
masses relatively close to the electroweak scale. Imposing the
new-physics threshold to be below the TeV scale, we have found several
minimal solutions for string-scale unification. All of them require at
least three new matter states.  It is remarkable that the addition of
three vector-like fermion doublets ($n_Q=3$) yields unification at the
string scale $\Lambda_S$ for
$(k_Y\,,k_2\,,k_3)=(13/3\,,1\,,2)\,$. These values are consistent with
the affine levels of an $SU(5) \times SU(5)$ string-GUT (see Table
\ref{tab:1}). In this case, the strong coupling constant at the $M_Z$
scale is $\alpha_s(M_Z) = 0.119$, with all the other electroweak input
data given at their central values.

The string landscape allows for a high-scale supersymmetry breaking. If
supersymmetry is broken at the string scale, most of its problems, such as fast
dimension-five proton decay, excessive flavor and $CP$ violation and stringent
constraints on the Higgs mass, are avoided. In this scenario, the Higgs boson
mass is predicted in the range $125~\text{GeV} \lesssim m_H \lesssim
170~\text{GeV}$, for the minimal string unification solutions presented here.

\ack D.E.C. would like to thank the kind hospitality of the organizers
  of the 8$^\mathrm{a}$ Hellenic South European School on Elementary
  Particle Physics, ``Quest for Unification: Theory Confronts
  Experiment'' (RTN meeting), 4 - 26 September 2005, Corfu, Greece.
  The work of D.E.C. and R.G.F. are supported by
  \emph{Funda\c{c}\~{a}o para a Ci\^{e}ncia e a Tecnologia} (FCT,
  Portugal) under the grants SFRH/BPD/1598/2000 and
  SFRH/BPD/1549/2000, respectively.

\end{document}